\documentclass[fleqn,12pt,twoside]{article}

\usepackage[headings]{espcrc1}

\readRCS
$Id: espcrc1.tex,v 1.2 2004/02/24 11:22:11 spepping Exp $
\usepackage{graphicx}
\usepackage[figuresright]{rotating}

\newcommand{\AmS}{{\protect\the\textfont2
  A\kern-.1667em\lower.5ex\hbox{M}\kern-.125emS}}

\title{Systematic Study of Identified Particle Production in PHENIX}

\author{M. Konno\address[MCSD]{Graduate School of Pure and Applied Sciences, 
        University of Tsukuba, Tsukuba, Ibaraki 305-8571, Japan} 
        for the PHENIX\thanks{For the full list of PHENIX authors and acknowledgments,
        see Appendix 'Collaborations' of this volume.} Collaboration}

\begin{document}

\maketitle


\begin{abstract}

Large enhancement of (anti)protons relative to pions has been
observed at intermediate $p_T$ $\sim$ 2-5 GeV/c in central Au+Au collisions at RHIC. 
To investigate the possible source of this baryon enhancement, 
we performed a systematic study of identified hadron spectra 
in Au+Au and Cu+Cu collisions at $\sqrt{s_{NN}}$ = 200 GeV, 
and Au+Au collisions at $\sqrt{s_{NN}}$ = 62.4 GeV.
The data set allows us to study the energy dependence and system size dependence
of the baryon enhancement.
We also compare the nuclear modification factors 
on hadron production in two different collision systems.

\end{abstract}


\section{Introduction}

One of the most remarkable observations in central heavy ion collisions at RHIC
is a large enhancement of baryons and antibaryons at intermediate $p_T$ $\sim$ 2-5 GeV/c, 
while neutral pions and inclusive charged hadrons are strongly suppressed 
compared to p+p collisions. \cite{phnx_pid,phnx_pi0,phnx_ch}. 
The (anti)proton to pion ratio is enhanced by almost a factor of 3 
when one compares most central Au+Au events to peripheral or p+p events.
In this $p_T$ region, fragmentation process dominates the particle production in p+p collisions. 
It is expected that the fragmentation is independent of the collision system, - hence 
the large baryon fraction observed at RHIC comes with a surprise.
This behavior is usually called ``Baryon anomaly at RHIC''.
By performing a control experiment - d+Au collisions, in which only cold nuclear matter is produced,
we found that the suppression of particle yields comes from not the initial state interactions, 
but the final state interactions (i.e. jet quenching) \cite{phnx_dAu}.
On the other hand, the observed enhancement is explained in some different ways:
(1) strong radial flow which pushes the heavier particles to larger $p_T$ \cite{model_hydro},
(2) recombination of shower quarks with quarks from the thermalized medium \cite{model_recomb_fries,model_recomb_hwa,model_recomb_greco}.
To investigate the origin of this anomaly, 
we use the several data sets from the past RHIC runs, 
including a new data set of Cu+Cu collisions taken in 2005 by PHENIX experiment.
The data set allows us to study the collision energy and system size dependences
of the hadron production.


\section{Data Analysis}

Data presented here includes Au+Au and Cu+Cu collisions
at $\sqrt{s_{NN}}$ = 200 GeV, and Au+Au collisions at $\sqrt{s_{NN}}$ = 62.4 GeV.
Events with vertex position along the beam axis within $|z|$ $<$ 30 cm
were triggered by the Beam-Beam Counters (BBC) located at $|\eta|$ = 3.0-3.9.
Using BBC, the triggered events are classified in collision centrality. 
Charged particles are reconstructed at mid-rapidity $|\eta|$ $<$ 0.35
using a drift chamber and multi-wire proportional chambers with pad readout. 
Particle identification is based on particle mass calculated from the measured momentum
and the velocity obtained from time-of-flight and path length along the trajectory.
The measurement uses the time-of-flight detector with the resolution of $\sigma$ $\sim$ 130 ps.
Corrections to the charged particle spectrum for geometrical acceptance, decay
in flight, reconstruction efficiency, and momentum resolution are determined 
using a single-particle GEANT Monte Carlo simulation.
Multiplicity-dependent corrections are evaluated by embedding simulated tracks into real events. 
Any feed-down corrections from weak decays are not applied to these results. 
The detailed analysis methods are found in \cite{phnx_pid}.


\section{Results}

\subsection{Baryon Enhancement - $p$/$\pi$ ratios at $\sqrt{s_{NN}}$ = 62.4 GeV}
\begin{figure}[h!]
\begin{center}
\includegraphics[height=6.5cm, width=15cm]{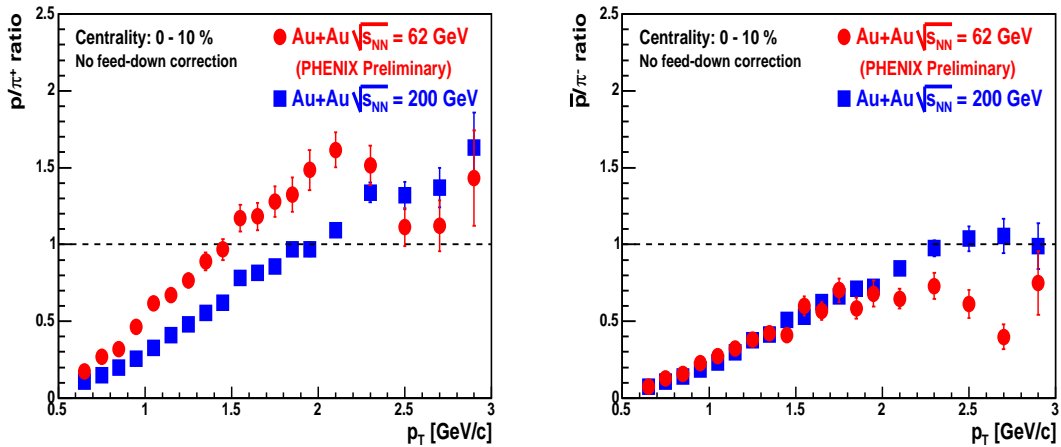}
\caption[]{p/$\pi^{+}$ (left) and $\overline{p}$/$\pi^{-}$ (right) ratios 
in central (0-10\%) Au+Au collisions at $\sqrt{s_{NN}}$ = 62.4 GeV and 200 GeV. 
Note that the feed-down corrections from weak decays are not applied.}
\label{fig:ppi_62_200}
\end{center}
\end{figure}

To study the excitation function of hadron production at the beam energy between SPS and RHIC, 
the data in $\sqrt{s_{NN}}$ = 62.4 GeV in Au+Au was taken. 
This lower energy data provides an important information on the baryon production and transport 
at mid-rapidity between SPS and RHIC. Figure~\ref{fig:ppi_62_200} shows the p/$\pi^{+}$ 
and  $\overline{p}$/$\pi^{-}$ ratios in central Au+Au collisons 
at 62.4 GeV and 200 GeV \cite{phnx_62_chujo}. 
Comparing to the 200 GeV data, the 62.4 GeV data shows 
a slightly larger proton contribution at intermediate $p_T$, 
while there is a less antiproton contribution.
The larger value of p/$\pi$ is explained by the larger difference 
between the slopes of spectra from fragmentation and recombination processes 
at 62.4 GeV than that at 200 GeV \cite{model_recomb_62}.

\newpage

\begin{figure}[h!]
\begin{minipage}[t]{75mm}
\includegraphics[height=6cm, width=7.5cm]{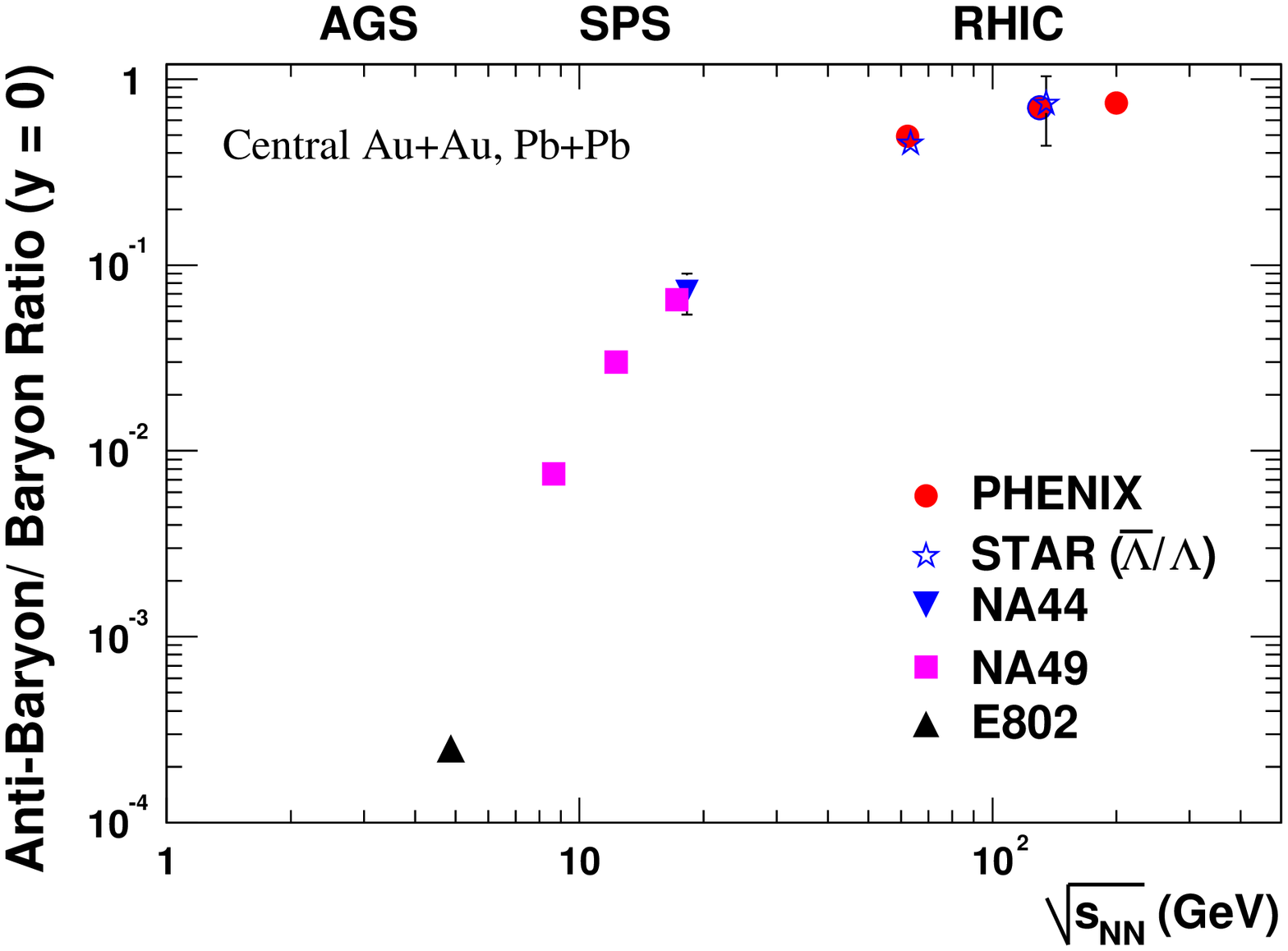}
\caption[]{Beam energy dependence of antibaryon-to-baryon ratios from AGS to RHIC.
The data point at 62.4 GeV from PHENIX is preliminary result.}
\label{fig:sqrts_pbarp}
\end{minipage}
\hspace{\fill}
\begin{minipage}[t]{75mm}
\includegraphics[height=6cm, width=7.5cm]{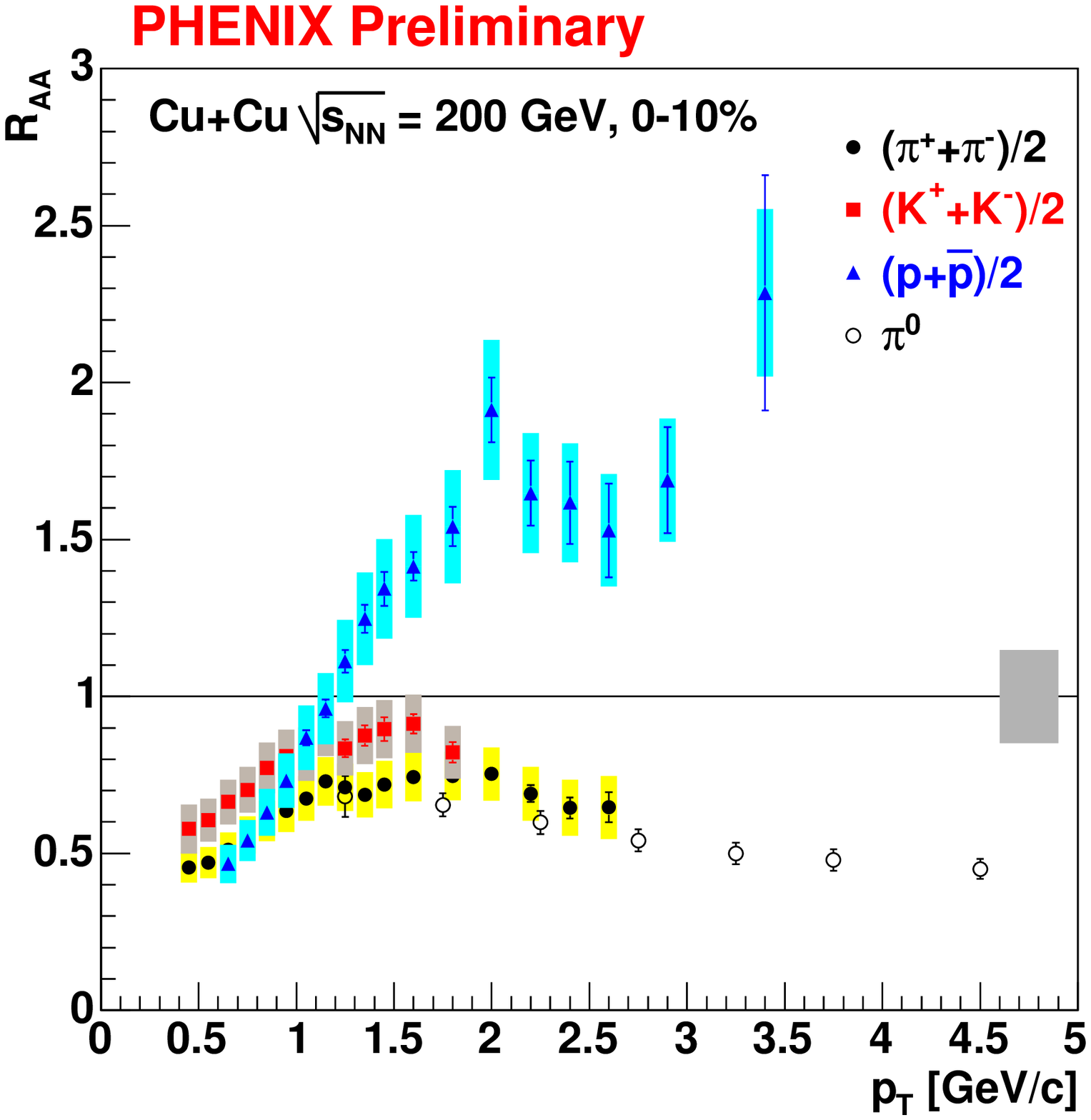}
\caption[]{Nuclear modification factors $R_{AA}$ for pions, kaons, and protons 
in central Cu+Cu ($N_{part}$ $\sim$ 98) collisions at $\sqrt{s_{NN}}$ = 200 GeV. 
}
\label{fig:raa_cucu200}
\end{minipage}
\end{figure}

Figure~\ref{fig:sqrts_pbarp} shows antibaryon-to-baryon ratios as a function of collision energy.
In 62.4 GeV Au+Au collisions, $\overline{p}$/$p$ ratio is about 0.5, independent of $p_T$,
and follows the smooth curve from SPS to RHIC.
It is consistent with the preliminary $\overline{\Lambda}$/$\Lambda$ ratio measured by STAR.
This indicates that the baryon chemical potential is larger at 62.4 GeV 
than that at 200 GeV in Au+Au collisions.

\subsection{$N_{part}$ Scaling - Cu+Cu vs. Au+Au}
~~

\subsubsection{Nuclear Modification Factors $R_{AA}$}
Figure \ref{fig:raa_cucu200} shows the nuclear modification factors $R_{AA}$ 
(ratio of the yields to the p+p yields scaled by the number of binary collisions) 
in central Cu+Cu collisions at $\sqrt{s_{NN}}$ = 200 GeV. 
When almost the same number of participants ($N_{part}$, presenting the system size) 
is chosen in Cu+Cu and Au+Au collisions, 
similar $R_{AA}$ trends in both collision systems can be seen
in terms of 
(1) magnitude of the enhancement / suppression 
(2) $p_T$ dependence
and (3) particle species dependences.
This could be called $N_{part}$-scaling of $R_{AA}$ 
between different collision systems at the same energy.\\


\subsubsection{$p$/$\pi$ Ratios}
\begin{figure}[h!]
\begin{center}
\includegraphics[height=6cm, width=7.5cm]{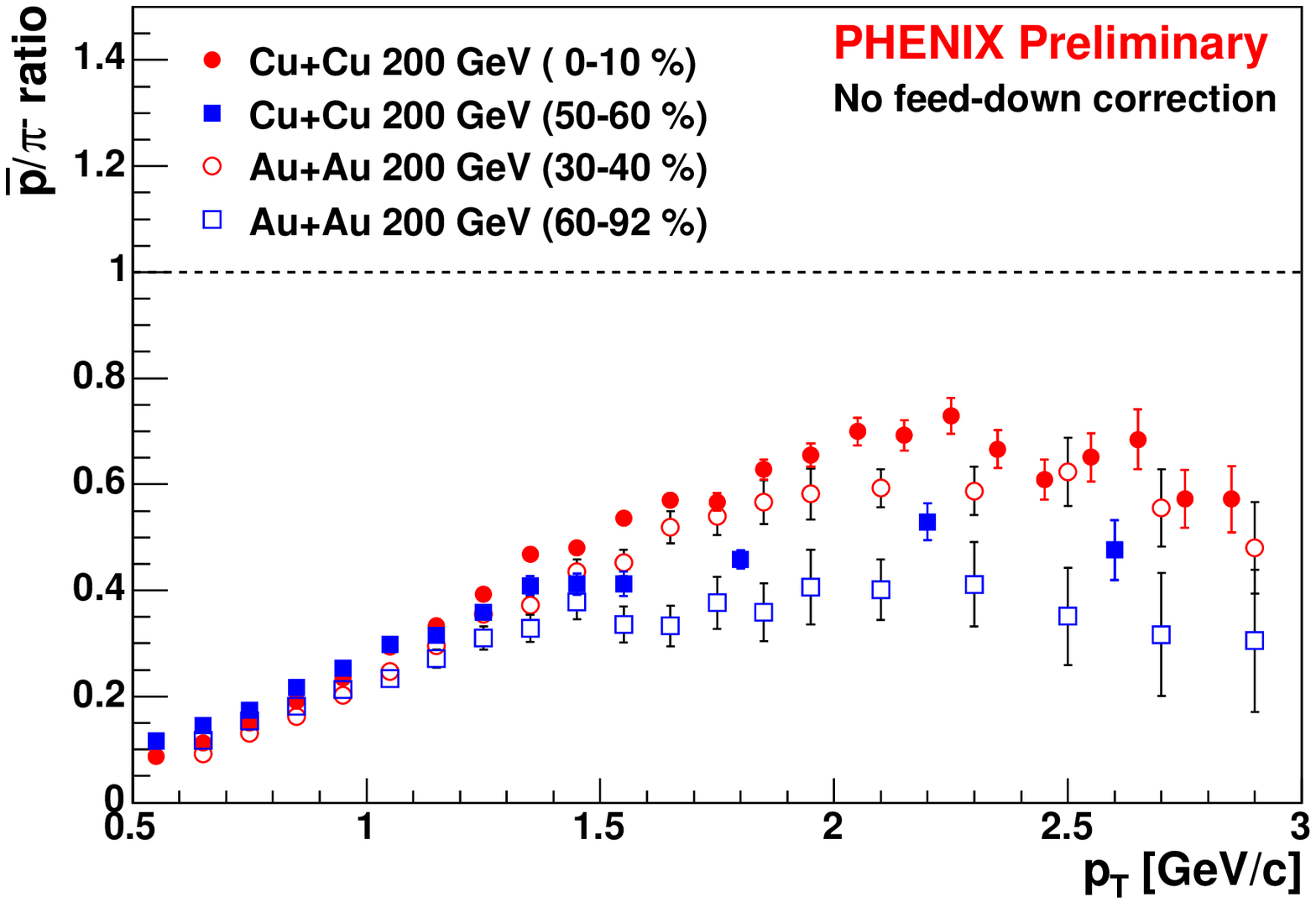}
\includegraphics[height=6cm, width=7.5cm]{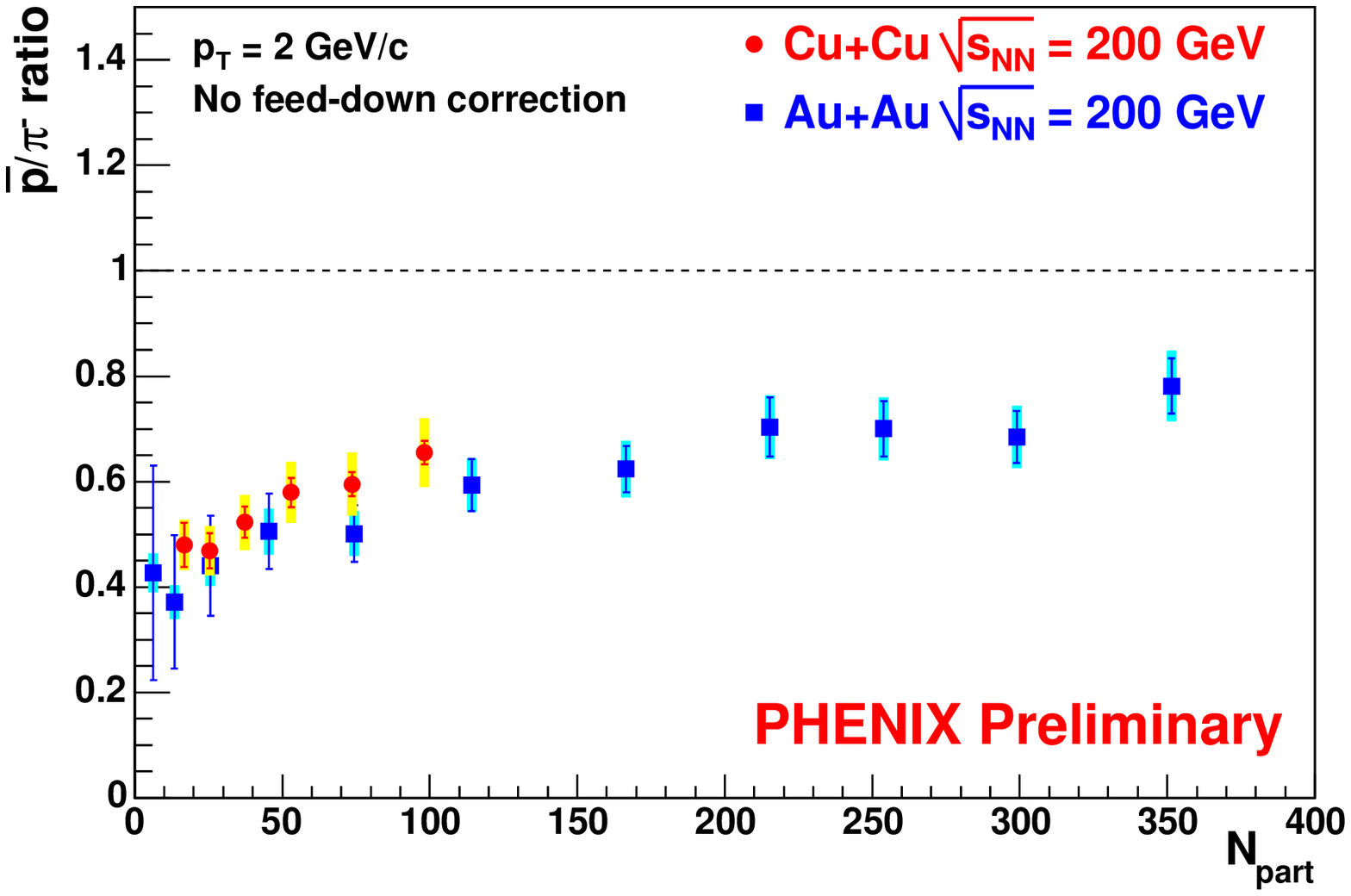}
\caption[]{$\overline{p}$/$\pi^{-}$ ratio 
as a function of $p_T$ (left) and as a function of $N_{part}$ at $p_T$ = 2 GeV/c (right)
in Cu+Cu and Au+Au collisons at $\sqrt{s_{NN}}$ = 200 GeV. 
Note that the feed-down corrections from weak decays are not applied.
}
\label{fig:ratio_ppi_vs_pt_npart}
\end{center}
\end{figure}

Figure \ref{fig:ratio_ppi_vs_pt_npart} shows the $\overline{p}$/$\pi^{-}$ ratio 
as a function of $p_T$ (left) and as a function of $N_{part}$ at $p_T$ = 2 GeV/c (right)
in Cu+Cu and Au+Au collisions at $\sqrt{s_{NN}}$ = 200 GeV. 
The $N_{part}$ is associated with centrality using a Glauber model calculation. 
The data shows similar $p_T$ and system-size ($N_{part}$) dependences, 
as seen in 200 GeV Au+Au,
despite the slight difference seen in the magnitude.
The $N_{part}$-scaling is also seen in p/$\pi^{+}$ and K/$\pi$ ratio.
This is another $N_{part}$-scaling on the particle ratios as well as $R_{AA}$.

\section{Summary}

Results on identified particle production in Cu+Cu and Au+Au collisions 
at $\sqrt{s_{NN}}$ = 200 GeV and Au+Au collisions at $\sqrt{s_{NN}}$ = 62.4 GeV 
are presented. 
The baryon enhancement at intermediate $p_T$ is observed in all collision systems.
The 62.4 GeV Au+Au data shows a slightly larger proton contribution at intermediate $p_T$ 
as seen in 200 GeV Au+Au, while there is a less antiproton contribution compared to 
the 200 GeV results.
These behaviors indicate 
that the hadron production at intermediate $p_T$ is 
due to multiple processes like fragmentation and recombination,
and that the baryon chemical potential is larger at 62.4 GeV than that at 200 GeV 
in Au+Au collisions.
The 200 GeV Cu+Cu data shows the $N_{part}$ scaling of p/$\pi$ and K/$\pi$ ratios
as seen in Au+Au data at the same energy.
The $N_{part}$ scaling is also seen in nuclear modification factor
in terms of the magnitude and particle species dependences.


\end{document}